\documentclass[a4paper,11pt]{article}

\usepackage{amsmath,amsfonts,amssymb,times,mathptmx,graphicx,amsthm}
\usepackage{placeins}
\usepackage[T1]{fontenc}
\usepackage[latin1]{inputenc}
\title{\bf Beyond the $\sqrt{\mathrm{N}}$-limit of the least squares resolution and the lucky model}
\author{Gregorio Landi$^a$\thanks{Corresponding
author. Gregorio.Landi@fi.infn.it}~,   Giovanni E. Landi$^b$\\
\\
\llap{$^a$} Dipartimento di Fisica e Astronomia,
Universita' di Firenze and INFN\\
Largo E. Fermi 2 (Arcetri) 50125, Firenze, Italy\\
\\
\llap{$^b$} ArchonVR S.a.g.l.,\\
Via Cisieri 3,
6900 Lugano, Switzerland.\\ \\
{\em April 29, 2019}}
\date{ }
\begin{document}
\maketitle 
\begin{abstract}
A very simple Gaussian model is used to illustrate a new fitting result: a linear growth of the resolution with  the number N of detecting layers.
This rule is well beyond the well-known rule proportional to $\sqrt{N}$  for the resolution of the usual fit.
The effect is obtained with the appropriate form  of the variance for each hit (measurement). The model reconstructs straight tracks with N parallel detecting layers,
the track direction is the selected parameter to test the resolution. The results of the Gaussian model are compared with  realistic simulations
of silicon microstrip detectors. These realistic simulations suggest an easy method to select the essential weights for the fit:
the lucky model. Preliminary results of the lucky model show an excellent reproduction of  the  linear growth of the resolution,
very similar to that given by realistic simulations.
\end{abstract}

Keywords: {\small Least squares method, Resolution,
Position Reconstruction, Center of Gravity,\newline\indent
Silicon Micro-strip Detectors, Lucky Model.}


\tableofcontents

\pagenumbering{arabic} \oddsidemargin 0cm  \evensidemargin 0cm

\section{Introduction}\indent

An essential source of information in high energy physics experiments is
the tracking of ionizing particles.
To accurately collect this information,  large arrays of particle detectors
(trackers) are installed in the experiments.
Among the types of particle detectors, silicon microstrip detectors are
frequently selected as tracker   components. A  silicon microstrip detector
has a very special surface treatment (strips) able to give a set of localized
signals (hit) to the readout system if a ionizing particle crosses the detector.
The algorithms for track reconstruction are essential completions of the
tracking systems and large efforts are dedicated to optimize their efficiency.
Our refs.~\cite{landi05,landi06} were dedicated to these improvements.
There we introduced  very special probability distributions, one for each hit,
calculated  for minimum ionizing particles (MIPs) crossing a microstrip detector.
These probability distributions are constructed around the form of the
hit positioning algorithms. For the hit positioning algorithm we used the
center of  gravity (COG), the easiest and most frequently employed algorithm
($ x_{gj}=\sum_{i=1}^j\  E_i \tau_i/\sum_{i=1}^j\ E_i $, where  $E_i$ is
the signal of the strip $i$, $\tau_i$ its position and $j$ the number of
accounted strips). Even if the COG is often indicated as a single algorithm,
it has different forms  and properties depending by the number $j$ of strip
signals inserted in the algorithm. Each COG form has very different analytical
and statistical properties and the mixture of different forms must be accurately
avoided. In the present developments, the two or three strip COG (COG$_2$ or
COG$_3$) are the only used forms. Their general expressions contain  ratios of
random variables that introduce Cauchy-like tails in their probability distributions.
Equation 3 of ref.~\cite{landi06}  illustrates this property for the COG$_2$ algorithm.
The long equations for the COG$_j$ probability distribution,
extended form of eq.5 of ref.~\cite{landi06}, are only a part of our
needs, to be useful in a fit, they must be completed with the functional
dependence from the MIP impact point. This dependence is inserted with the
functions of eq.3.2 of  ref.~\cite{landi05}, they were extracted from
the data of a test beam with the use of a theorem demonstrated in section
3.2 of ref.~\cite{landi05}. These functions are expressed with a Fourier series
with many terms (150-200). The essential deviation of our complete  probability
distributions from  Gaussian forms obliges the use of maximum likelihood method.
With such complex functions, the maximum-likelihood method requires many cares to
avoid the non-convergence of the search at the absolute maximum.
To improve the convergence  we developed a lower level fitting tool (called
schematic model) to be used as  initialization of the likelihood exploration.
The schematic model was built around effective variances extracted from our
complete probability distributions, cutting their Cauchy-like tails to avoid
divergences (eq.11 ref.~\cite{landi06}). The $\eta$-algorithm
gives the reconstructed hit positions (eq.8 of ref.~\cite{landi06}).
The definition of an effective variance
(weight) for each hit allows to determine the initial parameters of the
likelihood exploration with a weighted least squares. The first  application
of our complete method was the reconstructions of straight tracks
(ref.~\cite{landi05}), in this case the likelihood is a surface and the
convergence to the maximum can be easily followed. The results of the fits
showed drastic improvements of the reconstructed track parameters compared to standard fits.
The complete method is able to obtain excellent fits even in presence of large
outlayers, hence our method  eliminates  the outlayers from list of fitting problems.
For its simplified form, the schematic model is unable to handle the outlayers,
but it is very realistic in all the other aspects.
After acquiring confidence with straight tracks, we tested the reconstruction
of tracks of MIPs in a homogeneous magnetic field (ref.~\cite{landi06}).
Even the momentum reconstruction turns out much better than that of the standard fit.
In fig. 8 of ref.~\cite{landi06} we reported an interesting effect: an approximate
linear growth of the momentum resolution with the number $N$ of detecting layers.
This result is strikingly different from the textbook result that easily demonstrates a
growth in resolution as $\sqrt{\mathrm{N}}$ in least squares. Even if
$\sqrt{\mathrm{N}}$ refers to the fit of a constant, we will recall this type of rule
even for other fitted parameters where the rule is not so simple. The
aim of this work is the illustration of the generality of the linear growth
in resolution. For this, we study the fit of the track direction of
a MIP crossing a simpler tracker at orthogonal incidence. The tracker model is formed by $\mathrm{N}$
detecting layers of identical technology, without magnetic field and
exposed to a set of parallel tracks of high momentum MIPs (to neglect
the multiple scattering). To fix the ideas we suppose silicon micro-strip
detecting layers, but,  this one is not an essential condition. The simplicity
of the simulations gives a direct explanation of the results
of our complete fitting method.
Let us briefly recall the principal assumptions contained in the standard least
squares method.  Very often , identical variance for all measurements is
assumed~\cite{mathstat}. The identity of the variances is defined in statistics
as homoscedasticity. It introduces a drastic shortcut
in the least squares equations that become independent from the data variances. The opposite
of homoscedasticity is indicated as heteroscedasticity. In this case the identity of the
variances is abandoned. The equations for the least squares report the
weighted form (as in ref.~\cite{particle_group}), but, without
additional mathematical tools (as those we developed)  it is impossible to
consistently handle all the forms of  heteroscedasticity.
In any case, it is evident that the complexity introduced by heteroscedasticity
must be finalized to a gain of resolution.
To illustrate this point beyond the results of refs.~\cite{landi05,landi06} and
their long maximum likelihood searches, a simple Gaussian model will
be tested. The model easily shows the generation of the linear growth
with N of the resolution even with a very simple form of heteroscedasticity.
This Gaussian model is compared with a more realistic model
(the schematic model of refs.~\cite{landi05,landi06}).
We limit the schematic model to a single detector type, very
similar to silicon strip detectors largely used in running CERN-LHC
experiments~\cite{ALICE,ATLAS,CMSa}.
A preliminary introduction of a fast suboptimal tool (the lucky model) will
be discussed and compared with the schematic model. We limit our
discussion to the recipes to obtain these results. Mathematical details will be
published elsewhere.

\begin{figure} [h!]
\begin{center}
\includegraphics[scale=0.7]{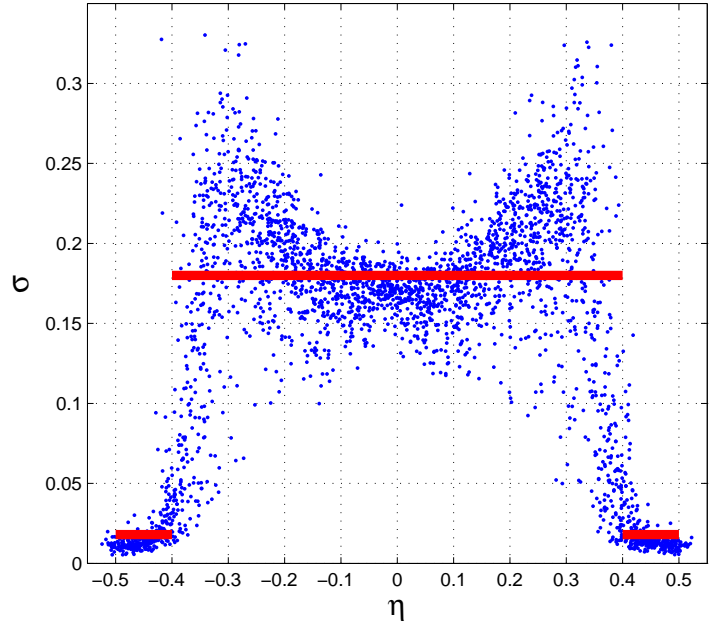}
\caption{\em The red lines are the values of $\sigma$ for the Gaussian model.
The blue dots are the effective $\sigma$ of the schematic model.
The dimensions are in strip width ($63 \mu \mathrm{m}$).
The $\eta$-algorithm gives the hit positions, with zero as the strip center.}
\label{fig:figure_1}
\end{center}
\end{figure}

\section{A simple Gaussian model and the linear growth}\indent

The model we explore is a Gaussian model with two different standard
deviations ($\sigma$), one of 0.18 (in unity of the strip width $ 63\, \mu m$)
with a probability of 80 $\%$ and the other of 0.018 with a probability of 20$\%$:
hard symmetry with a small "spontaneous" symmetry breaking.
The relation of this model with the schematic model of silicon micro-strips is
illustrated in fig.~\ref{fig:figure_1}. It represents a drastic simplification
of the models of refs.~\cite{landi05,landi06}. Experimental hints of the
border effect, evident in  fig.~\ref{fig:figure_1}  as a strong decrease
of the effective $\sigma$, are reported in ref.~\cite{CMSm} for gas ionization chambers.

Few lines of MATLAB~\cite{MATLAB} code suffice to produce this simulation, and they
could be a viable substitute of long mathematical developments. A large number of
Gaussian random numbers, with zero average and unity standard deviation,
are generated (with the MATLAB {\tt randn} function). Each one is multiplied
by one of the two standard deviations ($0.18$, $0.018$)
with the given relative probability ($80\%$, $20\%$).
The data are scrambled and recollected to simulate a set of parallel tracks crossing few detector layers.
This data collection simulates a set of tracks populating
a large portion of the tracker system with slightly non-parallel strips on different
layers (as it always is in real detectors). The detector layers are supposed parallel,
but for the non-parallelism of the strips, the hit positions of a track, relative to the
strip centers, seem to be uncorrelated. The properties of the binomial distribution 
produce the linear growth.  
Due to the translation invariance, the tracks can be expressed by a single
equation  $x_i=\beta+\gamma y_i$ with $\beta=0$  and $\gamma=0$ for
the orthogonal incidence of the set of tracks ($\beta$ is the impact point
of the track, $\gamma$ the direction, $y_i$
the positions of the detector layers and $x_i$ the hit position).
The distributions of  $\gamma_f$, the $\gamma$ value given by the fit, are
the object of our study. The $\gamma$ distribution is a Dirac $\delta$-function,
as it is usual to test the resolution of the fitted values $\gamma_f$.
The distance of first and the last detector layer is the length of
the PAMELA~\cite{vannu} tracker (445 $\mathrm{m m}$, 7063.5 in strip width),
but this is not essential.
Other "detector layers" are inserted symmetrically in this length.
Two different least squares fits are compared.
One uses identical $\sigma$ of each hit (we call this standard fit).
The other fit applies the appropriate $\sigma$-values to each hit. This second fit
shows a linear growth in the resolution (as far as a set of random variables can follow this rule). We generate 150,000 tracks for each configuration.

\begin{figure} [h!]
\begin{center}
\includegraphics[scale=0.65]{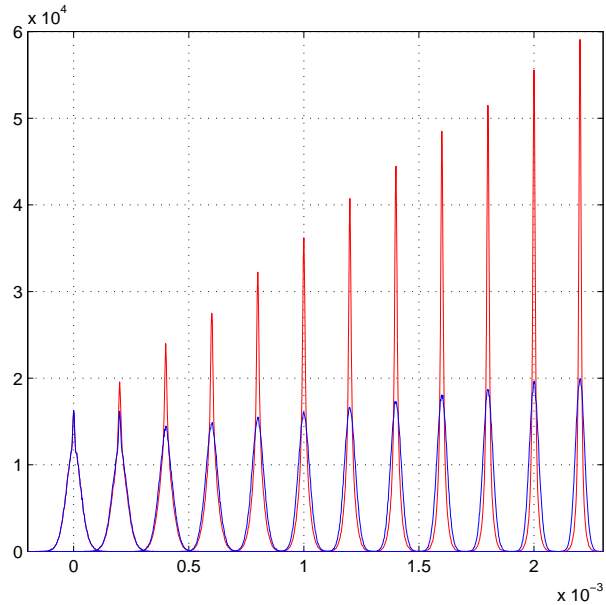}
\caption{\em The Gaussian model.
The blue lines are the $\gamma_f$-distributions of the standard fit.
The red lines are the $\gamma_f$-distributions of the $\sigma$-weighted
least squares, they show a linear growth in the resolution. The first distribution
is centered on zero, the others are shifted by N-2 identical steps.
}
\label{fig:figure_2}
\end{center}
\end{figure}

The results are reported in fig.~\ref{fig:figure_2} as
(empirical) probability density functions of $\gamma_f$. The parameter $\gamma_f$ is the tangent of a small
angle, a pure small number. To give a scale to the plots of $\gamma_f$, we could identify the tangent with
its argument and consider the horizontal scale in radians (rad) and the vertical scale as rad$^{-1}$. We will neglect all these details in the plots that
report probability density functions of pure number variables.

All the distributions we have to plot are centered around their fiducial value  $\gamma=0$ for each number N of layers.  For clarity of the plot, the  distributions with two layers is centered on zero.
The others with three, four, etc. up to thirteen layers are shifted by $\mathrm{N}-2$
identical steps to show better their increase. (This rule is extended to all similar plots.) Heteroscedasticity influences the first two
distributions (two and three layers) even in the standard fit. In fact, if the two hits of a track
have a narrow (Gaussian) position distributions (small $\sigma$), the $\gamma_f$ of the fitted tracks
is forced to be contained in a narrow distribution. It is curious that to reach the
height of these two distributions the standard fits require many additional layers.
We will consider the height of each distribution as an evaluation of the resolution of the corresponding
fit. Actually, the most common distributions have the maximum proportional to the inverse
of the full-width-at-half-maximum (FWHM). The FWHM is often considered a measure of the mean error
for non-Gaussian distributions and its inverse is the resolution. If the detector layers
are slightly different (as usual) the linear growth will show small distortions due to these differences.

\begin{figure} [h!]
\begin{center}
\includegraphics[scale=0.65]{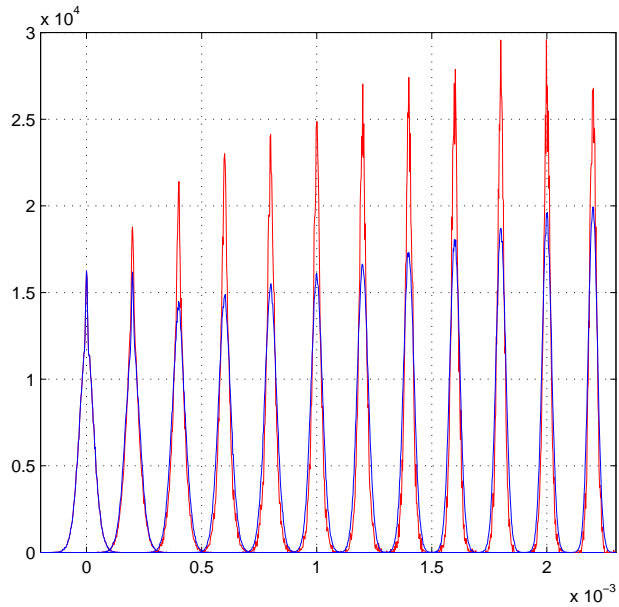}
\caption{\em The Gaussian model. Extraction of an approximate linear growth
from the standard least squares with the selection of low $\psi^2$-values. }
\label{fig:figure_3}
\end{center}
\end{figure}

An approximate linear growth can be extracted even from the standard least squares,
as illustrated in fig.~\ref{fig:figure_3}.

In fact, for a pure homoscedastic Gaussian model, it is a basic demonstration~\cite{mathstat} that the
fitted parameters are independent random variables (easily verified in these
simulations using a single average $\sigma$).
The presence of a small heteroscedasticity destroys this independence and
tends to couple the probability distribution of $\gamma_f$ with the values
of $\psi^2=\sum_i(x_i-\beta_f-\gamma_f y_i)^2$ for each track. The parameters
$\beta_f$ and $\gamma_f$ are those given by the fit of that track.
Among the lowest values of the $\psi^2$, good $\gamma_f$-values are more frequent to find.
The Gaussian model shows a partial linear increase of the $\gamma$ distributions selected
with $\psi^2$. We used the following empirical equation $\psi^2<0.08 (N/13)^2$ to account
for the $\psi^2$ increase with N.
Figure~\ref{fig:figure_3}  shows the result of the selection, this violation can be an
independent test of heteroscedasticity.

\section{The schematic model}\indent

Similar results can be obtained from our schematic model. This realistic model is called schematic
in refs.~\cite{landi05,landi06}. There it was used a first approximation of the complete model and as the starting point
for the maximum likelihood search. The quality of the fit produced by
the schematic model is not far from the complete model, the main differences are in the ability
of the complete model to find good results even in presence of the worst hits ({\em outliers}).
The calculations of the effective variance  of each hit is very time consuming, but after an initial (large)
set of $\sigma$ is obtained, the others can be quickly produced with interpolations
(high precisions are inessential). We always used the time consuming procedure.
Figure~\ref{fig:figure_1} shows a subset of effective $\sigma$
for this type of micro-strip detectors.
An interesting aspect of this scatter-plot is the very low values of $\sigma$ at the strip borders.
The positioning algorithm is the $\eta$-algorithm of ref.~\cite{belau}, briefly
summarized at pag.8 of ref.~\cite{landi06}. This algorithm is
essential to eliminate the large systematic errors of the two strip center of gravity (COG$_2$).
The forms of those systematic errors were  analytically described in ref.~\cite{landi01},
many years later than their full corrections.
References~\cite{landi03,landi04} added further refinements to the $\eta$-algorithm
and extended it to any type of center of gravity (COG) algorithm.

We underline
that the results, shown in the following, are impossible without the $\eta$-algorithm.
For example, the COG$_2$ algorithm degrades its (modest) results with this refined approach.
It is uncertain that the statistics has any relation with least squares fits
based on COG positioning algorithms. We proved in ref.~\cite{landi06}
that the elimination of any random noise (the statistics) does
not modify the probability distributions of the fit products based on the COG$_2$ positions.
Instead, the distributions of the fit results based on the $\eta$-algorithm rapidly converge
toward Dirac-$\delta$ functions, as it must be.

\begin{figure} [h!]
\begin{center}
\includegraphics[scale=0.7]{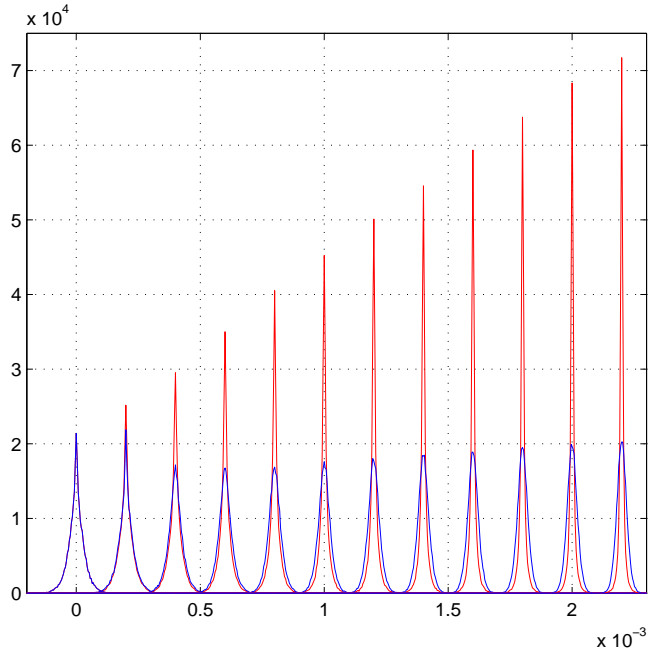}
\caption{\em Schematic model. The blue $\gamma_f$ distributions  are the
standard least squares fits. The red $\gamma_f$ distributions show the linear growth
and are produced with the effective $\sigma$ for each hit}
\label{fig:figure_4}
\end{center}
\end{figure}

Again, our aim will be the fit of straight tracks of equation $x_i=\beta+\gamma y_i$ with $\beta=0$ and
$\gamma=0$ incident orthogonally to a set of detector layers.
The properties of the simulated hits are modeled as far as possible on test-beam data~\cite{vannu},
as discussed in ref.~\cite{landi05}.
Even now, we compare the $\gamma_f$ distributions given by two types of least squares.
One fit uses the different $\sigma$ (the blue dots) of the scatter plot of
fig.~\ref{fig:figure_1} for the hit, and the
standard least squares that assumes identical variances for the hits. Figure~\ref{fig:figure_4}
shows these results.

\section{The lucky model}\indent

\begin{figure} [h!]
\begin{center}
\includegraphics[scale=0.65]{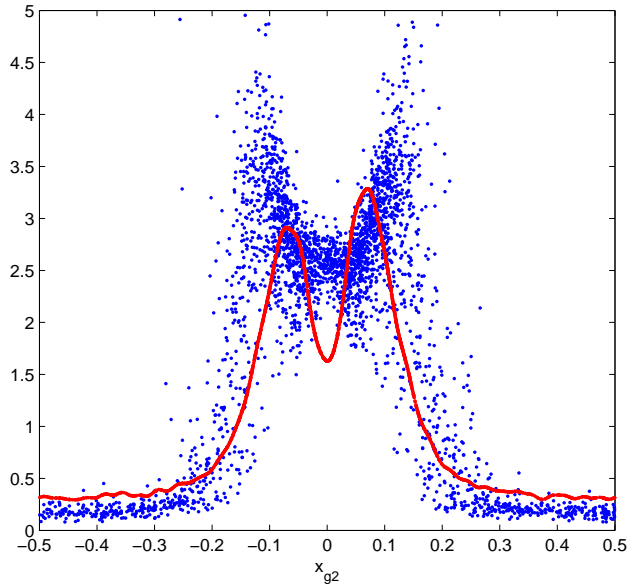}
\caption{\em Scatter plot of the effective $\sigma$ as function of the COG$_2$ ($x_{g2}$) scaled
to reach an approximate overlap with the COG$_2$ histogram. More precisely a merged
set of red dots corresponding to the blue dots. They are obtained as by-product of
the $\eta$-algorithm}
\label{fig:figure_5}
\end{center}
\end{figure}

In figure 1 of ref.~\cite{landi06}, a large asymmetry is evident in one of the two scatter plots.
Given our selection of an orthogonal incidence, symmetry is expected, but our long list of equations, required to arrive to the effective variances,
produce this asymmetry. The sole tentative explanation was the strong similarity of our scatter plots with the COG$_2$ histograms of the original data. One of the two histograms has the corresponding asymmetry~\cite{landi05}.
 The data were collected in
the test beam of   ref.~\cite{vannu}.
Our preliminary analysis did not find any simple relation (scaling factors)
able to reproduce the plots. The cut selection, we used in the definition of
the effective variances (eq.11 of ref.~\cite{landi06}),
were essentially based on an aesthetical criterion. We considered relevant the
reproduction of parts of the Gaussian features when they were present. Essentially
we tuned the cuts on a small subset of excellent hits to seek a comparison with our
complete probability distributions.
The scatter plots of ref.~\cite{landi05} were produced for the first time just
for their insertion in the publication. However, once we discarded the possibility
of a casual matching, a reasonable explanation could be found of this coincidence.
Thus, a more accurate analysis was undertaken. The use of the equations of ref.~\cite{landi06}
allow the construction a very approximate  demonstration
of the consistency of the trends of our scatter plots. A by-product of that is
a geometrical support to the resolution increase at the strip borders, just the origin of
the linear growth. All the details of this preliminary demonstration will
be published elsewhere, with the hope to obtain something better.
For example, simple functions that take into account
signal-to-noise ration of each hit.  Figure~\ref{fig:figure_5}
illustrates this very rough matching of the $\sigma$ scatter plot and the COG$_2$ histogram.
The $\sigma$-values are scaled to produce an approximate overlap with the COG$_2$ histogram.
More precisely, we plot the normalized histogram divided by the bin size and interpolated with
a Fourier series, the red line is a merged set of dots produced by the interpolation.
One red dot for each blue dot.

Here we recall the intuitive explanation given in ref.~\cite{landi05} about this
similarity. The effective $\sigma$
estimates the ranges of the possible impact values converging to the same COG$_2$ value.
Hence, larger $\sigma$ gives higher COG$_2$ probability and lower $\sigma$
gives lower COG$_2$ probability. Inverting this statement, it looks reasonable that hits
with the lower COG$_2$ probability have lower effective $\sigma$ and hits in the higher COG$_2$
probability have higher effective $\sigma$. If these assumptions would be successful
we have a very economic strategy to implement heteroscedasticity in the track fitting with
pieces of information that were well hidden just in front of us. But, without
the hints of the scatter plots of ref.~\cite{landi05}, they could
remain hidden. The COG$_2$ probabilities are very easy to obtain from the corresponding
histograms or as a by-product of the $\eta$-algorithm~\cite{landi03}.
Further details are required: the scaling factors to render compatible the
COG$_2$ histograms and effective $\sigma$ scatter plots.
In general, these scaling factors must be calculated. But,
for identical detectors, as in our case, an identical factor is required
and it becomes irrelevant for the linearity of the
(weighted) least squares equations. Hence we can attempt to directly use
the amplitude of histogram as rough effective $\sigma$
and observe the effects in the fit.

\subsection{The linear growth in the lucky model}\indent

\begin{figure} [h!]
\begin{center}
\includegraphics[scale=0.7]{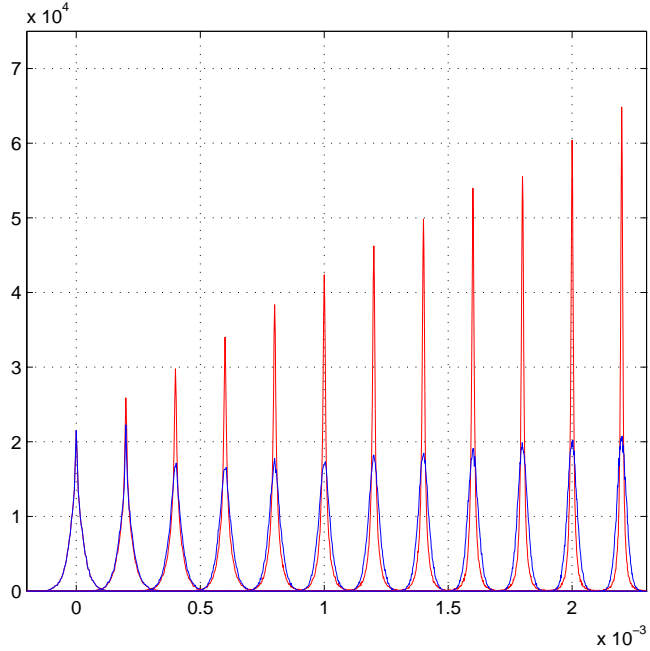}
\caption{\em The lucky model. The blue distributions are the
standard least squares. The red distributions are
the $\gamma_f$ given by the lucky model, very near
to those of the schematic model.  }
\label{fig:figure_6}
\end{center}
\end{figure}

The "lucky" results are illustrated in fig.~\ref{fig:figure_6} for this
"lucky" model. This figure is produced as fig.~\ref{fig:figure_4}
with the sole difference given  by the use of the rough
effective $\sigma$ in the weighted least squares.

The maximum of fig.~\ref{fig:figure_6} is
around $6.5\,10^4$ and fig.~\ref{fig:figure_4} is around $7.2\,10^4$ (a difference of $10\%$):
the results of the fits are excellent. The robustness
of the heteroscedasticity is evident. Large variations of the details of the
parameters (probability distributions) of the models
do not modify the fit results. For example, the use of the
effective variances, illustrated
with the red line in fig.~\ref{fig:figure_1}, with the $\eta$-positioning gives
$\gamma_f$-distributions identical to the complete schematic model.
This result is easy to justify, the weights inserted in the least squares are $\sigma^{-2}$,
and the higher $\sigma$-values  are all compressed near to zero, and their
differences, compared to the constant value of fig.~\ref{fig:figure_1}, become negligible.
Instead, the lower $\sigma$-values are very similar in the two
models and they dominate the fits, exactly as for the lucky model.
Another connection is contained in the lucky model; the functional form
of $\sigma^{-1}$ as a function of $\eta$ is well-known function introduced
with a theorem in ref.~\cite{landi01} and illustrated in ref.~\cite{landi03}.
This function is the average shape of the MIP signal collected by the strip,
hence higher signal density gives better hits.

We even test  the lucky model  approximation with the three strip
COG (COG$_3$) histogram and its corresponding $\eta$ positioning
($\eta_3$). Even here, the linear growth is clearly evident but the
maximum is around $1/2$ of that of fig.~\ref{fig:figure_4}. The strip added
to the COG$_2$, to compose the COG$_3$, is almost always pure
noise  at this orthogonal incidence~\cite{landi03}.
In addition, the COG$_3$ has discontinuities
at the strip borders~\cite{landi01}, here very small and masked by
the noise, but just in the region with the best $\sigma$. Thus this
lower result is not unexpected, and it is consistent with
our full three strips schematic model.
Similar tests were performed on the low noise
floating strip side.

\section{Hints for an experimental verification}\indent

Some of the results presented here could be verified with a test beam.
The beam divergence should be less than $10^{-5}$ radians, probably
not easy to obtain. In the simulations we used a very simplified (for
the computer) approach: the detector-layer configurations were generated
each time dividing symmetrically the allowed interval from the first and last
layer. This corresponds to a different experimental set-up for any
layer number. A more realistic set-up could be composed with a fixed number of parallel
detector layers, (normal silicon micro-strips) orthogonal to
the beam (of high momentum to limit the multiple scattering). An effective increase of the
layer number can be produced varying those inserted in the fit. Excellent precision in the
tracker alignment parameters and a very small beam divergence are required for a
direct test of the linear growth. The $\eta$-algorithm
is essential. At orthogonal incidence and without magnetic field the $\eta$-corrections
are very small or negligible. In any case for parallel tracks the corrections are
identical and their neglect implies a parallel translation of the tracks. The $\psi^2$
and the lucky model can be used to test heteroscedasticity.

If the beam divergence is large compared to the  $\gamma_f$-resolution of the fit,
the presence of the linear growth in resolution can be observed in the fluctuation of the difference
of the $\gamma_f$ fitted with all the available layers and the $\gamma_f$ fitted with the
elimination of a layer each time in the same track. The linear growth produces a
parabolic increase of these differences. Instead, the standard fit produces a small increase in the last
couple of differences.

\section{Conclusions}\indent

Simple simulations produce linear growths in the fit resolution with the number $N$ of
detector layers, similar to those of ref.~\cite{landi06} for the momentum reconstruction.
Here, sets of parallel straight tracks  are used in the fits, and the
test parameter is the direction $\gamma$ of the tracks. This two-parameter fit
is easier than that for the momentum. The Gaussian model easily produces  the
linear growth of the fit resolution. The model is so simple, for its essential
mathematics, that it can be completed with  few lines of
MATLAB code and its results are almost identical to our very complicated
schematic model. The addition of the physics of the
detectors is a heavy task. The evident similarity of the effective $\sigma$
scatter-plots with the histograms of the two strips center of gravity
triggered a more accurate analysis of its origin. This analysis was sufficiently
convincing to try a weighted least squares fit with weights extracted from the center of
gravity histogram (the lucky model). The result of this test is excellent with
a drastic increase of the fit resolution and the sought linear growth.
The differences of the lucky model compared to the schematic model are around
$10\,\%$, a negligible price compared to the enormous simplification in the the extraction
of the weights. It is evident that these are very preliminary results, and
further tests are essential before a systematic use of the model. Very synthetic indications
are given for a experimental verification of the model.



\end{document}